\documentclass[aps,prb,twocolumn,showpacs,tightenlines]{revtex4}
\usepackage{epsfig}
\begin{document}

\bibliographystyle{prsty}

\def \cT {{\cal T}}
\def \cI {{\cal I}}
\def \cf {{\cal f}}
\def \cG {{\cal G}}
\def \cD {{\cal D}}
\def \cU {{\cal U}}
\def \cV {{\cal V}}
\def \cF {{\cal F}}
\def \cT {{\cal T}}
\def \cH {{\cal H}}
\def \cA {{\cal A}}
\def \cL {{\cal L}}
\def \cR {{\cal R}}
\def \cN {{\cal N}}
\def \cC {{\cal C}}
\def \cS {{\cal S}}
\def \cP {{\cal P}}
\def \cE {{\cal E}}
\def \cM {{\cal M}}

\title{Temperature Dependence of Resistivity of $Sr_2CoMoO_{6-\delta}$ Films}

\author
{C. L. Yuan$^1$, Z. Y. Zeng$^{1,2}$, Y. Zhu$^1$, P. P.
Ong$^1$\cite{byline}, Z. X. Shen$^1$, C. K. Ong$^3$} \affiliation
{$^1$Department of Physics, National University of Singapore, 2
Science Drive 3,
Singapore, 117542\\
$^2$ Department of Physics, Hunan Normal University, Changsha
410081, China\\ $^3$ Centre for Superconducting and Magnetic
Materials, Department of Physics, National University of
Singapore, 2 Science Drive 3, Singapore, 117542}

\date{\today}

\begin{abstract}
We investigate the temperature dependence of the resistivity and
magnetoresistance of a polycrystalline $Sr_2CoMoO_{6-\delta}$ film
deposited on (100)-$SrTiO_3$ substrate prepared by the pulsed
laser deposition method.  X-ray diffraction, Raman and
magnetoresistance results demonstrate clearly the coexistence of a
ferromagnetic metallic and an antiferromagnetic (or paramagnetic)
insulating domain. Percolative transition between these two phases
as the temperature varies, which is believed to induce a
metal-insulator transition at around $T_C$, has been directly
observed in our measurements of the temperature dependence of the
sample resistivity.  Thus we have provided new direct evidence
that a phase separation scenario also exists in the ordered
double-perovskite structure materials.

\end{abstract}

\pacs{72.15.Gd, 73.40.Gk, 75.50.Cc}

\maketitle

Since the observation of most significant colossal
magnetoresistance (CMR) effect close to the magnetic transition
temperatures, there has been an intense search for compounds with
magnetic transition temperatures substantially higher than the
$T_C$ ($\sim 200-350 K$) in manganites.\cite{Helmolt, Neumeier,
Schifferer} Recently, it has been reported that the
transition-metal oxides with the ordered double-perovskite
structure, $A_2MMoO_6$, ($A$ is a rare-earth metal and $M$ is a
transition metal), exhibits a pronounced negative CMR at lower
magnetic fields and higher temperatures compared to the doped
manganites.\cite{Kobayashi1,Borges,Kobayashi2,Viola} The reason
for this improved MR property in these compounds at a relatively
higher temperature arises primarily from the fact that they have a
surprisingly high magnetic transition temperature ($Sr_2FeMoO_6 ~
415 K$\cite{Kobayashi1} ; $Sr_2CoMoO_6 ~ 350 K$\cite{Viola})
compared to manganites. The largest MR response is expected close
to the magnetic transition temperature $T_C$.  The combination of
large MR effect and high $T_C$ value makes this family of
perovskites promising for practical applications.

 Up to present, many researchers have
reported the structural, magnetic, and electrical properties of
the double-perovskite transition-metal oxides, especially for the
Fe-based compounds. To the best of our knowledge, however, the
behavior of resistivity at high temperature especially near $T_C$
is less addressed and the temperature dependence of the
grain-boundary resistivity is not very well
understood.\cite{Niebieskikwiat} In this work we investigate the
temperature dependence of the resistivity and
magnetoresistance(MR) of a polycrystalline $Sr_2CoMoO_{6-\delta}$
film, especially its resistivity behavior at high temperatures.

 Experimental and theoretical
 investigations\cite{Uehara,Fath,Mereo1, Mereo2,Varelogiannis,Hotta,Mayr,Dagotto}
suggest that due to the tendencies toward phase separation, the
ground state of manganite models will not be a homogeneous state,
but an inhomogeneous one, typically involving both ferromagnetic
(FM) metallic and antiferromagnetic (AFM) charge and orbital
ordered insulating domains.\cite{Dagotto} The strong tendencies of
phase mixing originate from: (1) phase separation between
different-density phases arising from specific electronic
structures, and (2) disorder-induced phase separation with
percolative characteristics between equal-density phases driven by
disorder near first-order metal-insulator transitions. It is also
argued\cite{Dagotto} that the interesting mixed-phase tendencies
and percolation should be present in a large variety of other
compounds as well. In this paper we report the finding of actual
evidence of  percolative mixed-phase characteristics in the
temperature dependence of the resistivity of the polycrystalline
$Sr_2CoMoO_{6-\delta}$ films appearing as a metal-insulator
transition peak in the resistivity-temperature curve. This is the
first time that such a mixed-phase phenomenology has been observed
to appear in the ordered double-perovskite structures materials.

A Thin film of $Sr_2CoMoO_{6-\delta}$ of thickness  700 $nm$ was
deposited on a (100) $SrTiO_3$ (STO) substrate by pulsed laser
deposition. The substrate temperature during deposition was kept
at 500 $^oC$ in high vacuum ($10-7$ Torr). The crystal structure
and phase purity of the samples were examined by X-ray diffraction
using Cu $Ka$ radiation. The micro-Raman spectrum was measured in
the backscattering geometry using an ISA Jobin-Yvon-Spex T64000
Raman spectrometer with an Olympus microscope attachment. The
514.5 nm line of an argon-ion laser was used as the excitation
source. The electrical resistivity  $\rho$ was measured using the
standard four-probe technique. The magnetoresistivity ratio, MR,
is defined as MR$(\%)=\Delta \rho/\rho_0 \times 100\%$,   where
$\Delta \rho=\rho_0-\rho_H$, $\rho_H$ and $\rho_0$ are  the
resistivities in an applied magnetic field and in zero field,
respectively.

Fig. 1 depicts the X-ray diffraction pattern for the deposited
$Sr_2CoMoO_{6-\delta}$ thin films on STO at room temperature. It
shows that the film was polycrystalline, and a minor amount of
$SrMoO_4$ was also detected. The peak, labeled ``$\star$",
corresponds to the peak of the $SrMoO_4$ impurity phase. All the
other peaks corresponding to the main phase of
$Sr_2CoMoO_{6-\delta}$  can be indexed according to the space
group $I4/mmm$.

\begin{figure}
\epsfig{file=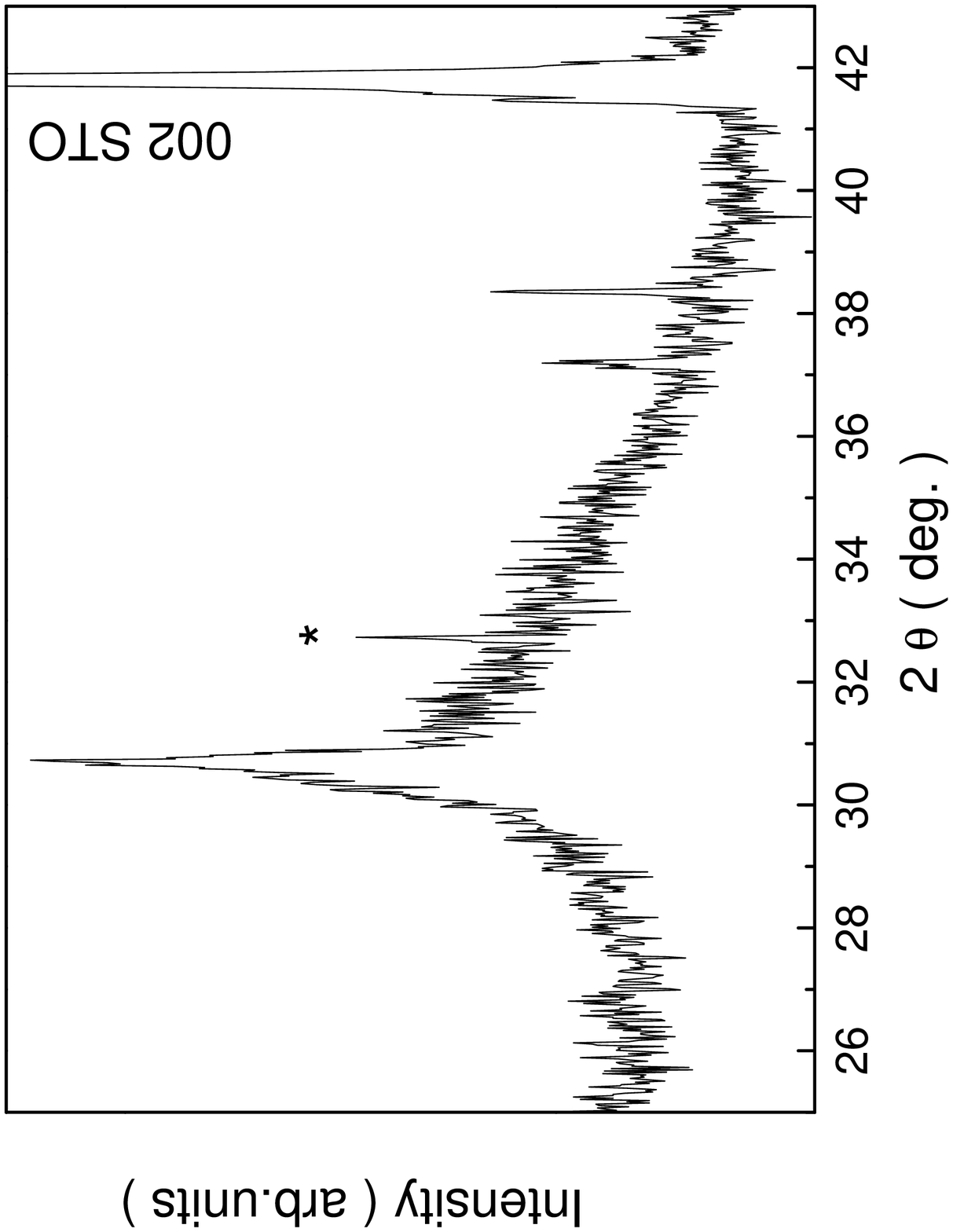, width=8.5cm} \caption{X-ray diffraction
patterns of polycrystalline $Sr_2CoMoO_{6-\delta}$
   thin film. The peak pertaining to the impurity $SrMoO_4$ is indicated by ``$\star$".}
\end{figure}

Fig. 2 shows the Raman spectra of the sample at room temperature.
The peak at $882.3$  $cm^{-1}$, labeled ``$\star$", corresponds to
the strongest Raman line of the $SrMoO_4$ impurity phase. The
presence of the $SrMoO_4$ impurity in the sample  reinforces the
conclusion obtained in the XRD analysis in Fig. 1.

\begin{figure}
 \epsfig{file=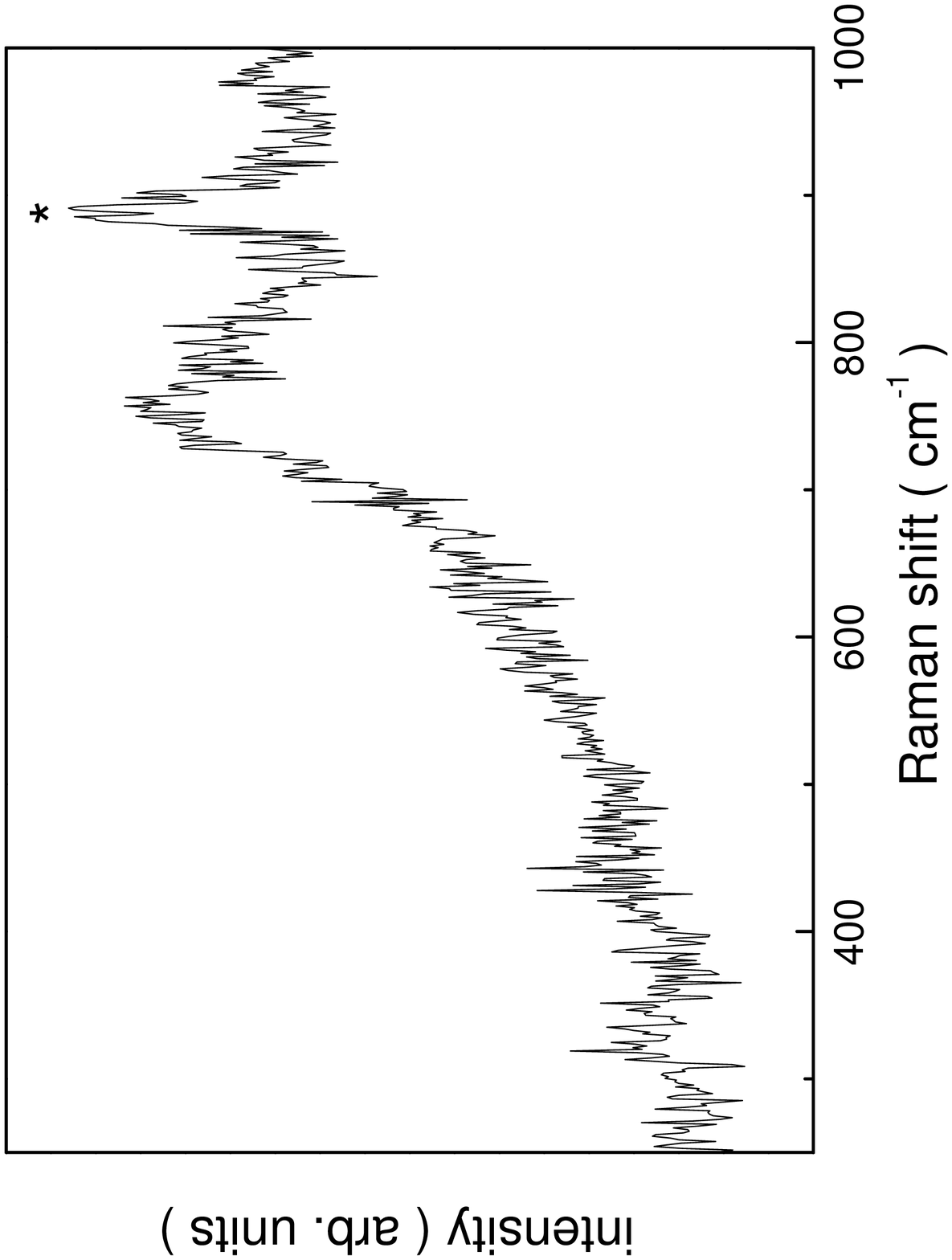, width=8.5cm}
  \caption{Raman spectra of polycrystalline $Sr_2CoMoO_{6-\delta}$ thin film at room temperature.
  The peak pertaining to the impurity $SrMoO_4$ is indicated by ``$\star$".}
  \end{figure}

 Fig. 3 shows
the temperature dependence of the resistivities at zero field and
at $1T$ respectively for the sample in the temperature range from
$80-300K$. At the lower temperature range,  The resistivity
decreases exponentially with temperature, showing a
semiconducting-like or thermally activated behavior. The
temperature dependence of the magnetoresistivity ratio
MR$(\%)=\Delta \rho/\rho_0 \times 100\%$ ($\Delta
\rho=\rho_0-\rho_H$ ) at 1T is presented in the insert of Fig. 3.
The maMR decays as the temperature increases, comparable to that
found in the other double perovskite
$Sr_2FeMoO_6$.\cite{Kobayashi1} The MR is dominated by the grain
boundary tunneling and demonstrates the existence of ferromagnetic
metallic clusters. The increasing value of MR as the temperature
is lowered is attributed to the weak spin thermal fluctuation at
low temperatures.

\begin{figure}
\epsfig{file=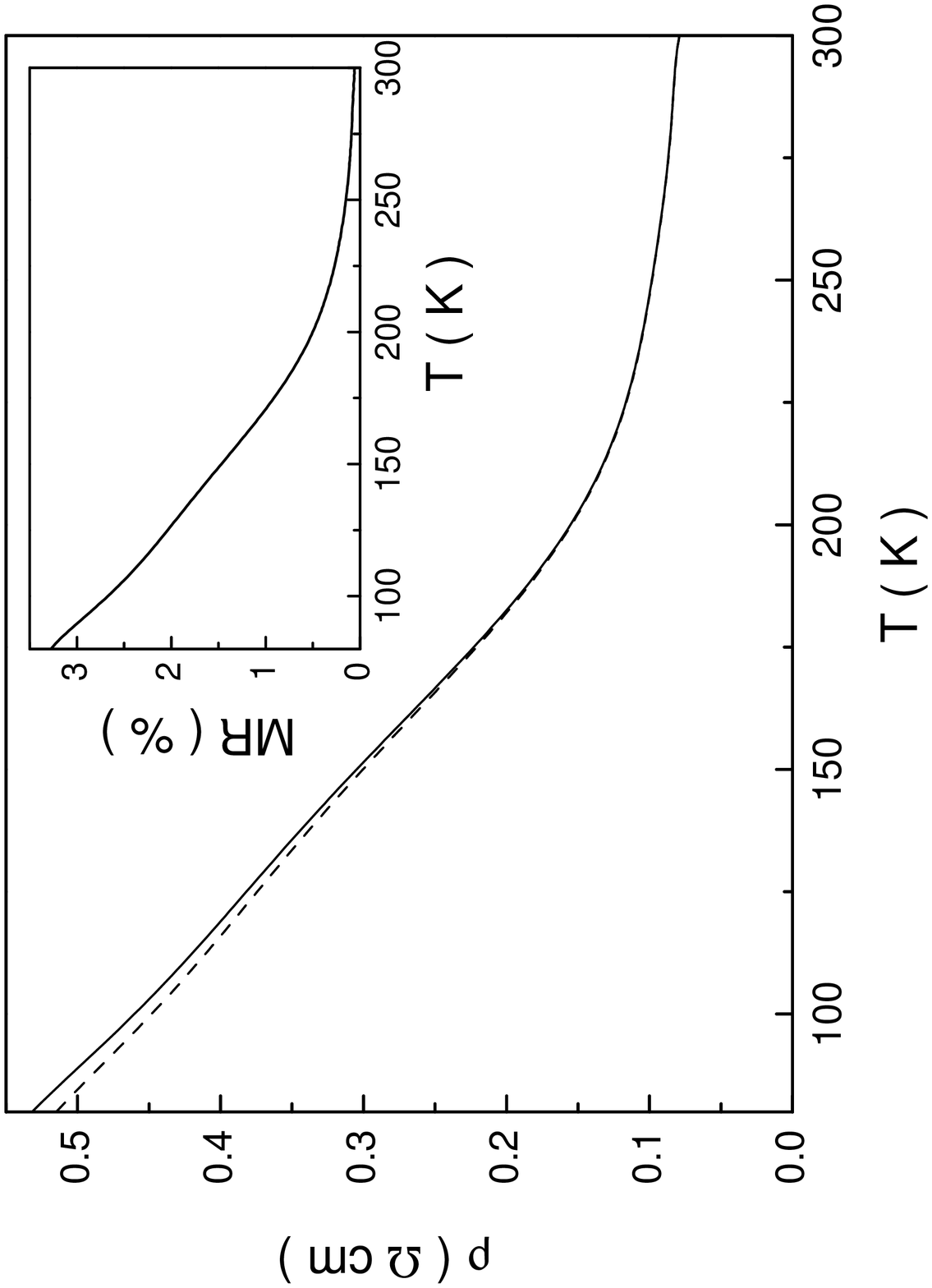, width=8.5cm}
  \caption{Temperature dependence of resistivity for polycrystalline $Sr_2CoMoO_{6-\delta}$
  thin film at zero field (solid line) and at $1T$ (dash line) at the temperature
   range from $80$ to $300 K$. The insert is temperature dependence of the magnetoresistivity
   ratio MR$(\%)=(\rho_0-\rho_H)/\rho_0 \times 100\%$
   for polycrystalline $Sr_2CoMoO_{6-\delta}$ thin film at $H=1T$
at the temperature range from $80$ to $300 K$.}
  \end{figure}

Fig. 4 shows the temperature dependence of the resistivity of the
$Sr_2CoMoO_{6-\delta}$ film kept in vacuum ($10^{-6}$ Torr) in the
extended temperature range from $80$ to $500 K$. The arrow
indicates the Curie temperature ($T_C$) determined by Viola et
al.\cite{Viola} At very low temperatures $Sr_2CoMo_{6}$ is known
to behave as a paramagnetic insulator,\cite{Viola,Itoh} which may
be ascribed to the absence of the $4d$ electrons in the hexavalent
$Mo^{6+}$($4d^0$) ions and the divalent nature of the valence of
the $Co$ ions.\cite{Sleight} Magnetic measurements suggest the
presence of ferromagnetic domains characterized by $T_C$ embedded
in an antiferromagnetic matrix (AFM) characterized by Niel
temperature $T_N$.\cite{Viola} Contrary to the $Sr_2FeMoO_6$
structure, the formation of ferromagnetic clusters is not due to
the double exchange mechanism\cite{Moritomo} but rather to a kind
of super-exchange interactions between $Co^{2+}$ and $Mo^{5+}$
moments. Based on a consideration of the small number of itinerant
electrons introduced from the oxygen vacancies, $Mo^{5+}$ cations
should be randomly isolated and so only isolated ferromagnetic
clusters can be formed.\cite{Viola} Therefore
$Sr_2CoMoO_{6-\delta}$ can be viewed as a canted-spin system or
one containing inhomogeneous ferromagnetic clusters embedded in an
paramagnetic matrix (or AFM matrix  if the temperature is below
$T_N$). At low temperatures the resistivity exhibits a
semiconductor-like or thermally activated behavior. The
resistivity of the $Sr_2CoMoO_{6-\delta}$ film is about two orders
of magnitude smaller than that of the bulk $Sr_2CoMo_{6}$,
implying that a higher density of charge carriers (electrons) is
involved in the conduction process. However, this resistivity is
still too high compared to that in a metallic state. This
indicates that the resistivity of the sample arises from the
carrier scattering at the ferromagnetic grain
boundaries.\cite{Kobayashi1} This observation is supported by the
inverse temperature dependence of the measured MR shown in the
insert of Fig. $3$.

\begin{figure}
\epsfig{file=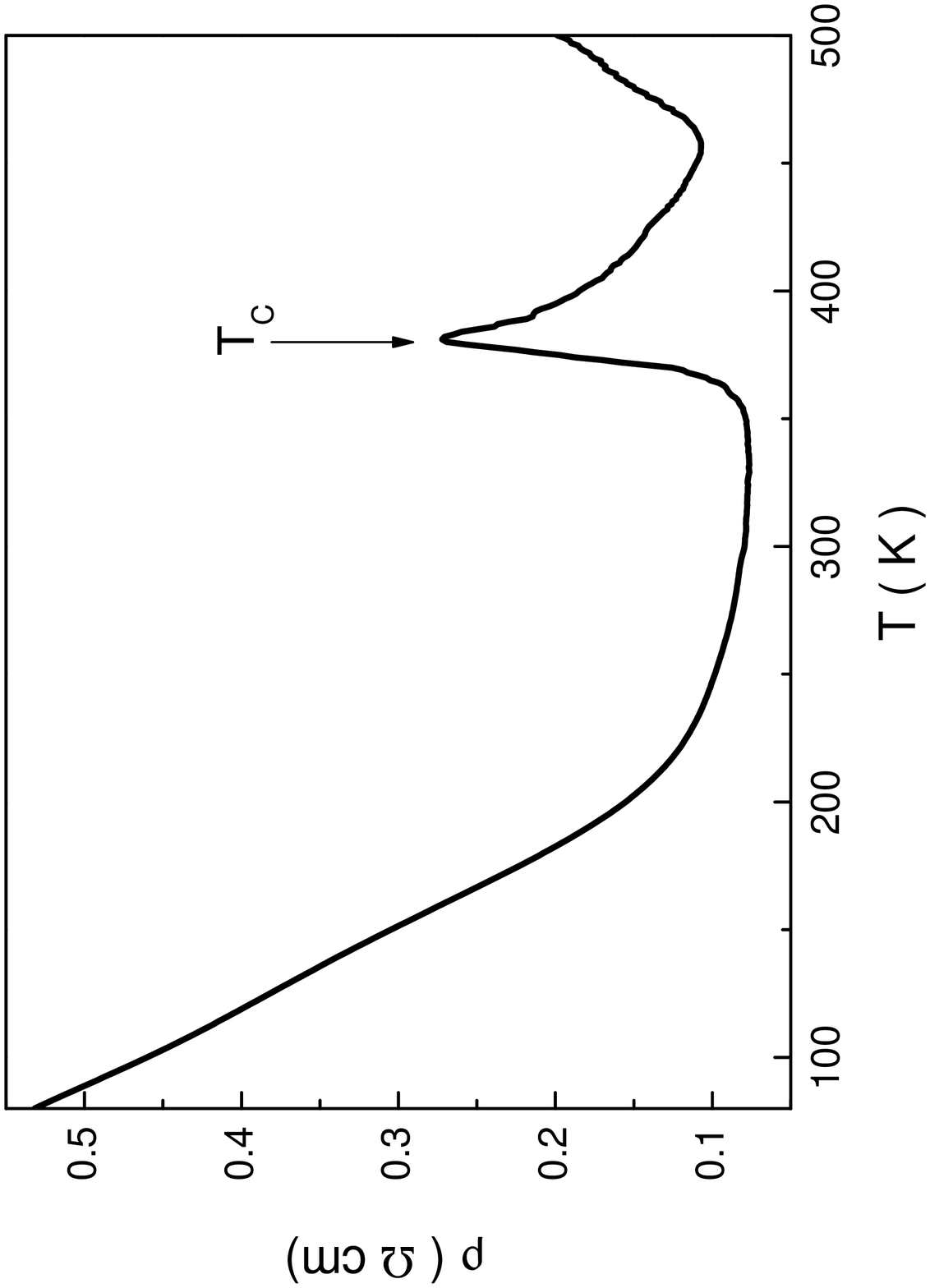, width=8.5cm}
  \caption{Temperature dependence of resistivity for polycrystalline
  $Sr_2CoMoO_{6-\delta}$ thin film at the temperature range from $80$ to $500 K$.}
  \end{figure}

A conspicuous change in the resistivity $\rho$ is observed at
around $T_C$. In the low-temperature ($T<T_C$) ferromagnetic
phase, a metallic behavior ($\rho$ increases with temperature) is
observed, while a semiconductor behavior is observed in the
paramagnetic phase ($T>T_C$ ). The semiconducting behavior at $T$
above $T_C$ could be due to the thermal spin fluctuation of the
ferromagnetic domains and the randomness of the boundary
scattering. With further increase in temperature significantly
beyond $T_C$, the resistivity shows a linear increase, which may
be ascribed to the enhanced phonon scattering of the carriers and
other localization effects.

The temperature dependence of the resistivity of
$Sr_2CoMoO_{6-\delta}$ can be well understood from the phase
separation scenario involving percolative transport through the
ferromagnetic clusters in a background AFM
matrix.\cite{Uehara,Fath,Mereo1,Dagotto} Starting from a regime
where the ferromagnetic clusters are formed dynamically (spin
fluctuation) at a high temperature $T^*$ above $T_C$, the metallic
state can be achieved if the temperature is decreased and the size
of clusters grows gradually until eventually percolative transport
through the ferromagnetic domains becomes possible. At this
temperature, the carriers can move over a long distance and the
metallic state is reached.\cite{Uehara,Fath,Mereo1,Dagotto}
Alternatively, one can image the metallic fraction to drop with
increasing temperature, which is a reasonable assumption since the
metallic portion of the sample originates from the ferromagnetic
(FM) arrangement of spins that improves conduction. Such a spin
arrangement deteriorates as the temperature increases. Moreover it
is reasonable to imagine that the size of FM clusters decreases as
the temperature grows. Then, a pattern of FM clusters that had
been connected at low temperature (leading to a metallic behavior)
may become disconnected at higher temperatures. The direct
observation of the metal-insulator transition peak appearing in
our temperature dependence of the sample resistivity represents
evidence of the onset of percolative transition, and hence the
existence of the phase separation scenario in the
$Sr_2CoMoO_{6-\delta}$ film. The decrease in the metallic portion
with increasing temperature was confirmed by a recent theoretical
investigation on the random field Ising model\cite{Mayr} based on
the random resistor network model.\cite{Kirkpatrick} However,
quantitative determination of the $T^*$ at which the ferromagnetic
clusters are formed is difficult and needs further investigation.

In summary, a polycrystalline $Sr_2CoMoO_{6-\delta}$ film was
fabricated and the behavior of its resistivity at high
temperature, especially around $T_C$, has been investigated. The
sample can be viewed as a typical mixed-phase system with
ferromagnetic metallic clusters embedded in the AFM insulating
matrix. With increasing temperature, its magnetoresistance
decreases until the room temperature, the sample resistivity
exhibits a metal-insulator transition peak near $T_C$, which is
attributed to the percolative transition between the FM metallic
and AFM insulating phases.  It provides the first experimental
evidence that the phase separation scenario also exists in the
transition-metal oxides with the ordered double-perovskite
structure.

 C. L. Yuan, Y. Zhu and P. P. Ong was supported under NUS Research Grant No. $R-144-000-011-012$ and $R-144-000-064-112$.
 Z. Y. Zeng  was supported under NUS Research Grant No. R-144-000-082-112.

\end{document}